\documentclass[doublecol,linenumbers]{epl2} 
\usepackage{graphicx}
\usepackage{dcolumn}
\usepackage{bm}
\usepackage{graphicx,amssymb,amsmath,amsthm,amsfonts,epsfig}

\title{Surface Area Products for Kerr-Taub-NUT Space-time }

\author{Parthapratim Pradhan\footnote{pppradhan77@gmail.com}\inst{1}}

\institute{                    
  \inst{1} Department of Physics, Vivekananda Satabarshiki Mahavidyalaya,
West Midnapur, West Bengal 721513, India \\
}

\abstract{
We examine properties of the inner and outer horizon thermodynamics of
Taub-NUT (Newman-Unti-Tamburino) and Kerr-Taub-NUT (KTN) black hole (BH) in
four dimensional \emph{Lorentzian geometry}. We compare and contrasted these
properties with the properties of Reissner Nordstr{\o}m (RN) BH
and Kerr BH. We focus on ``area product'', ``entropy product''
, ``irreducible mass product'' of the event horizon and Cauchy horizons. 
Due to mass-dependence, we speculate that these
products have no beautiful quantization feature. Nor does it has
any universal property. We further observe that
the \emph{First law} of BH thermodynamics and \emph {Smarr-Gibbs-Duhem }
relations do not hold for Taub-NUT (TN) and KTN BH in
Lorentzian regime. The failure of these aforementioned features
are due to presence of the non-trivial NUT charge which makes the space-time to be
asymptotically non-flat, in contrast with  RN BH and Kerr BH. The another
reason of the failure is that Lorentzian TN and
Lorentzian KTN geometry contains \emph{Dirac-Misner type
singularity}, which is a manifestation of a non-trivial topological twist
of the manifold. The black  hole \emph{ mass formula} and
\emph{Christodoulou-Ruffini mass formula}  for TN and
KTN BHs are also computed. This thermodynamic product formulae gives 
us further understanding to the nature of BH entropy (inner and outer) at 
the microscopic level.
}

\begin{document}

\maketitle

\section{Introduction}
Perhaps the most remarkable and beautiful analogy between the four laws of
BH physics and the laws of thermodynamics were first discovered in
the early 1970s \cite{cd,cr,pr,hk,bk,bk1,bc,bcw,hk1,hk3}. Particularly, the
Bekenstein-Hawking area-entropy relation which is described by the
well known formula:
\begin{eqnarray}
{\cal S}_{+} &=& \frac{{\cal A}_{+}}{4}
\end{eqnarray}
where, ${\cal S}_{+}$ is the Bekenstein-Hawking entropy (in units in which $G=\hbar=c=k=1$) and
 ${\cal A}_{+}$ is the area of the event horizon (${\mathcal H}^{+}$).

Another striking mathematical correspondence was established by Hawking's discovery
\cite{hk} that BHs radiate as perfect black bodies at
\begin{eqnarray}
T_{+} &=& \frac{\kappa_{+}}{2 \pi}
\end{eqnarray}
where $T_{+}$ is the Hawking temperature computed at the ${\mathcal H}^{+}$ and
$\kappa_{+}$ denotes the surface gravity of the BH computed at the ${\mathcal H}^{+}$.

It is a well-known fact that certain BH possesses inner horizon (${\mathcal H}^{-}$) or
Cauchy horizon (CH) in addition to the outer horizon or
event horizon (EH). While the above geometric quantity holds for ${\mathcal H}^{+}$,
there might have indication that these relations also hold for ${\mathcal H}^{-}$.
That means the Bekenstein-Hawking entropy  for ${\mathcal H}^{-}$ is given by
\begin{eqnarray}
{\cal S}_{-} &=& \frac{{\cal A}_{-}}{4}
\end{eqnarray}
where ${\cal A}_{-}$ is the area of the ${\mathcal H}^{-}$
and the Hawking temperature for ${\mathcal H}^{-}$ is
\begin{eqnarray}
T_{-} &=& \frac{\kappa_{-}}{2 \pi}
\end{eqnarray}
where $\kappa_{-}$ denotes the surface gravity of the BH
computed at the ${\mathcal H}^{-}$.
This idea first came from the fact that the  product of the EH
and the CH's  area of any asymptotically flat
BH admitting a smooth extremal limit, seems to depend only on the
quantized charge, quantized angular momentum  and is independent of the ADM (Arnowitt-Deser-Misner) mass
of the space-time \cite{larsen} and see also \cite{ansorg1,cvetic,castro,visser1,det,chen,pp14,hl,kntn}.

Specifically, in case of axisymmetric Einstein-Maxwell gravity, the product   of the
${\mathcal H}^{+}$ area and ${\mathcal H}^{-}$
areas  obey the following formula:
\begin{eqnarray}
{\cal A}_{+} {\cal A}_{-} &=& (8\pi)^2\left(J^2+\frac{Q^4}{4}\right) ~.\label{prKN}
\end{eqnarray}
which is remarkably  independent of the ADM (Arnowitt-Deser-Misner)  mass of the space-time.
Therefore it does depend on the quantized  charge $Q$, and quantized angular momentum $J$  of
the BH respectively \cite{ansorg1,visser1}. When $J=0$, we get the area 
product formula for RN BH:
\begin{eqnarray}
{\cal A}_{+} {\cal A}_{-} &=& (4\pi Q^2)^2 ~.\label{prrn}
\end{eqnarray}
which is indeed quantized and independent of ADM mass.

On the other hand, in multi-dimensional string theory/M-theory and followed by a holographic principle 
suggests that the product of Killing horizon areas for certain multi-horizon BHs are also 
in fact independent of the ADM mass. For asymptotically flat BPS (Bogomol'ni-Prasad-Sommerfield) 
BH in four and higher dimension, the areas of the inner and outer horizons are of the form 
${\cal A}_{\pm} =8\pi \left(\sqrt{N_{1}}\pm\sqrt{N_{2}} \right)$
and the product 
\begin{eqnarray}
{\cal A}_{+} {\cal A}_{-} 
=\left(8\pi \right)^2 N , \,\,\, N\in {\mathbb{N}},
N_{1}\in {\mathbb{N}}, N_{2} \in {\mathbb{N}}  ~.\label{ppl}
\end{eqnarray}
should therefore quantized in integral multiples of $(8\pi)^2$. Where the integers
$N_{1}$ and $N_{2}$ could be viewed as the excitation numbers of the left and right moving modes of a 
weakly coupled 2-dimensional conformal field theory (CFT) \cite{cvetic} and see also 
\cite{larsen,castro,det,chen}. $N_{1}$ and $N_{2}$ depend explicitly on all the BH 
parameters. Therefore the product ${\cal A}_{+} {\cal A}_{-}$ is also be an integer \cite{larsen,cvetic,castro} 
and from Eq. (\ref{prKN}) we find the product ${\cal A}_{+} {\cal A}_{-}$ is indeed quantized, and 
intriguingly, it is expressed solely in terms of the quantized charge and quantized angular momentum.  
Comparing Eq. (\ref{prrn}) and Eq. (\ref{ppl}), one can obtain the relation between
universality and quantization:
\begin{eqnarray}
\prod_{i} Q_{i} &=& \prod_{i} N_{i}
\end{eqnarray}
Where $i=1,2,3,4$,  $Q_{i}$ is the charges occur  only in quantized units and  $N_{i}$ is the 
integers. This can be treated as a relation between universality and quantization. Simply, for 
RN BH all charges are equal $Q_{i}=Q$. Curir \cite{crnvc} first showed that the first 
law of BH thermodynamics holds on the CH as well as EH:
\begin{eqnarray}
d{\cal M} &=& T_{\pm} d{\cal S}_{\pm} +\Omega_{\pm} dJ ~. \label{curi}
\end{eqnarray}
The same author also  calculated the ``area product'' and ``area sum'' of Kerr BH for the interpretation 
of the spin entropy  of the inner horizon. That is
\begin{eqnarray}
{\cal A}_{+} {\cal A}_{-} &=&  64 \pi^2 J^2 ~.\label{prkerr}
\end{eqnarray}
which is indeed quantized and independent of the mass,
and
\begin{eqnarray}
{\cal A}_{+} + {\cal A}_{-} &=&  16 \pi {\cal M}^2 ~.\label{skerr}
\end{eqnarray}
She also showed that the sum of entropy of both the horizons (${\mathcal H}^{\pm}$) are equal to  the entropy
of a Schwarzschild BH \cite{crgrg}. That is
\begin{eqnarray}
{\cal S}_{+} + {\cal S}_{-} &=&  4 \pi {\cal M}^2 ~.\label{sch}
\end{eqnarray}

In a very recent work \cite{wang}, the authors have investigated  an entropy sum relation  of  BHs in
four dimensional  RN-AdS and Kerr-AdS space-time  back-ground.  They showed that
the entropy sum and area sum depends on the cosmological parameter and also does not
depend on the ADM mass (${\cal M}$) and charge ($Q$) parameters.

It is true that  the CH has the property of being a null surface of
infinite blue-shift, while the  EH is an infinite red-shift surface \cite{sch}. It is
also true that the inner horizon of both the RN and Kerr BH's is highly
unstable due to the exterior perturbation. That means, the  energy-momentum tensor associated with
various massive and massless test fields have diverging value on the CH \cite{ps,chht}. 
In an another work, Poisson and Israel proved that when a RN BH is perturbed its inner horizon 
becomes a singularity of infinite space-time curvature which is known as mass inflation 
singularity \cite{pi}. 
For an external observer to observes any event that occurs  at ${\mathcal H}^{-}$,
he/she should follow the path as described by S. Chandrasekhar in Fig. (14), Page-212 by using maximal 
analytical extension i.e. Carter-Penrose diagram in case of RN spacetime \cite{sch}.

Thus in contrast with the outer horizon, there might be a relevance of the inner horizon in the
BH thermodynamics. Therefore the CH takes now an important position in BH
thermodynamics to explain the Bekenstein-Hawking entropy and different thermodynamic
products. This is why we study the following  thermodynamic products like area product,
entropy product and irreducible mass product of the EH and CHs.
We consider here Lorentzian TN BH and Lorentzian  KTN BH. We show whether the first law of BH
thermodynamics does hold for both Lorentzian  TN BH and Lorentzian  KTN BH. This is a crucial
point to investigate this properties due to the inclusion of the NUT parameter \cite{ntu} or gravito-magnetic
mass. A NUT charge can be defined as in terms of the field strength $F_{ab}$:
\begin{eqnarray}
n=-\frac{1}{4\pi} \int _{\cal S} {\bf F},
\end{eqnarray}
where ${\cal S}$ is any topological 2-sphere in the orbit space \cite{hunter98}.
We also compute the Smarr's mass formula \cite{smarr} and Christodoulou-Ruffini mass
formula for the above mentioned BH. This is the main purpose of this paper.

The plan of the paper is as follows.
In Sec. (\ref{tnt}), we describe various thermodynamic products for TN BH. In this
Sec., there are two sub-section. In first sub-section, we derive the mass formula for TN
BH. In second sub-section, we derive the Christodoulou-Ruffini mass formula for TN BH.
In Sec. (\ref{kerrt}), we discuss  various thermodynamic products for KTN BH.
In the first sub-section, we describe the Smarr's mass formula for KTN BH. In the
second sub-section, we derive the  Christodoulou-Ruffini mass formula for KTN BH.
Finally, we conclude in Sec. (\ref{dis}).

\section{\label{tnt}  Area Product and Area sum  of Lorentzian TN BH:}

To give a warm up,  we first consider the simple Lorentzian TN BH, it is a 
stationary, spherically-symmetric vacuum solution of Einstein equations with the NUT 
parameter $(n)$. This NUT charge or dual mass has an intrinsic feature in Einstein's 
general relativity and which is the gravitational analogue of a magnetic monopole in 
Maxwell's electrodynamics \cite{nouri}. The presence of the NUT parameter in the space-time
destroy its asymptotic structure making it, in contrast to the RN spacetime, asymptotically
non-flat.

Thus the metric is given by \cite{mistaub,kruskal}
$$
ds^2 = -U(r) \, \left(dt+2n\cos\theta d\phi\right)^2+
$$
\begin{eqnarray}
\frac{dr^2}{U(r)}+\left(r^2+n^2\right) \left(d\theta^2
+\sin^2\theta d\phi^2 \right) ~.\label{tnn}
\end{eqnarray}
$$
U(r) = 1-\frac{2({\cal M}r+n^2)}{r^2+n^2}
$$
Here, ${\cal M}$ represents the gravito-electric mass or ADM mass  and $n$ represents
the gravito-magnetic mass or dual mass or magnetic mass of the space-time. The idea
of dual mass  could be found in \cite{sen}. It is evident that there are two type of
singularities are present in the metric (\ref{tnn}). One type occurs at $U(r)=0$ which
give us the Killing horizons or BH horizons:
\begin{eqnarray}
r_{\pm}= {\cal M}\pm\sqrt{{\cal M}^2+n^2}\,\, \mbox{and}\,\,  r_{+}> r_{-}
\end{eqnarray}
Here, $r_{+}$ is called EH (${\cal H}^+$) or outer horizon and $r_{-}$  is called
CH  or inner horizon. The other type of singularity occurs at $\theta=0$
and $\theta=\pi$,  where the determinant of the metric component vanishes. Misner \cite{misner}
showed that in order to remove the apparent singularities at $\theta=0$ and $\theta=\pi$ , $t$
must be identified modulo $8\pi n$. Provided that $r^2+n^2 \neq 2({\cal M}r+n^2)$.
It should be noted that the NUT parameter actually  measures deviation from the asymptotic
flatness at infinity which could be manifested in the off-diagonal components of the metric and
this is happening due to presence of the Dirac-Misner type of singularities.

The area of both the horizons should be
\begin{eqnarray}
{\cal A}_{\pm} &=& \int^{2\pi}_0\int^\pi_0\sqrt{g_{\theta\theta}g_{\phi\phi}} d\theta d\phi \\
               &=& 8\pi\left[({\cal M}^2+n^2)\pm {\cal M}\sqrt{{\cal M}^2+n^2}\right]
~.\label{tn}
\end{eqnarray}
The product of areas yield 
\begin{eqnarray}
{\cal A}_{+} {\cal A}_{-} &=& (8\pi n)^2\left({\cal M}^2+n^2\right) ~.\label{pdtn}
\end{eqnarray}\
It indicates that the product is dependent on both mass parameter and NUT parameter. Thus it is 
not universal and not quantized. Here the quantization meaning  is that mass could not be 
discretized where as charge could be discretized. So when the area product of EH and Cauchy
horizons containing mass parameter obviously it should not be quantized. This is the relation between 
universality and quantization. Thus the conjecture ``area product is universal''
does not hold for non-asymptotic TN space-time.

Now the area sum could be obtained as,
\begin{eqnarray}
{\cal A}_{+}+ {\cal A}_{-} &=& 16\pi\left({\cal M}^2+n^2\right)~.\label{sumtn}
\end{eqnarray}
It implies that area sum does depend on the mass, and is therefore not universal.

The BH entropy of ${\cal H}^\pm$ is given by
\begin{eqnarray}
{\cal S}_{\pm} &=& 2\pi \left({\cal M}r_{\pm}+n^2 \right)  ~.\label{etptn}
\end{eqnarray}
The entropy product yields:
\begin{eqnarray}
{\cal S}_{+} {\cal S}_{-} &=& (2\pi n)^2\left({\cal M}^2+n^2\right) ~.\label{penttn}
\end{eqnarray}
and the entropy sum is given by 
\begin{eqnarray}
{\cal S}_{+}+ {\cal S}_{-} &=& 4\pi \left({\cal M}^2+n^2\right) ~.\label{ntrtn}
\end{eqnarray}
It indicates that entropy product and entropy sum are not universal.

The surface gravity computed at the  ${\cal H}^\pm$ is
\begin{eqnarray}
{\kappa}_{\pm} &=& \frac{r_{\pm}-r_{\mp}}{ 4\left({\cal M}r_{\pm}+n^2 \right)} \,\, \mbox{and}\,\,
\kappa_{+}> \kappa_{-} ~.\label{sgtn}
\end{eqnarray}
The BH temperature of ${\cal H}^\pm$  is given by
\begin{eqnarray}
T_{\pm} &=& \frac{r_{\pm}-r_{\mp}}{ 8 \pi \left({\cal M}r_{\pm}+n^2 \right)} ~.\label{tmtn}
\end{eqnarray}
It should be noted that $T_{+} > T_{-}$. In the limit ${\cal M}=0$, one obtains the result of 
massless TN BH.


\subsection{ The Mass Formula for  TN Spacetime:}

Smarr \cite{smarr} first showed that mass can be expressed as a function of area, angular momentum and
charge for Kerr-Newman BH. On the other hand,  Hawking \cite{bcw} pointed out that the BH
surface area can never decrease. Therefore the BH surface area is indeed a constant
quantity over the ${\cal H}^\pm$. For TN spacetime, it is given by Eq. (\ref{tn}). Inverting
this Eq. one can obtain for TN BH, the mass as a function of area of both the
horizons (${\cal H}^\pm$) and NUT parameter $n$:
\begin{eqnarray}
{\cal M}^2 &=&  \frac{1}{({\cal A}_{\pm}-4\pi n^2)} \left[ \frac{{\cal A}_{\pm}^2}{16\pi}
-n^2({\cal A}_{\pm}-4\pi n^2)\right]~.\label{masstn}
\end{eqnarray}
Let us calculate the mass differential $d{\cal M}$ to define the two physical quantities
in terms of inner and outer horizon.
\begin{eqnarray}
d{\cal M} &=& \Gamma_{\pm} d{\cal A}_{\pm} +\Phi_{\pm}^{n}dn
~. \label{dmtn}
\end{eqnarray}
where
\begin{eqnarray}
\Gamma_{\pm} &=&  \frac{{\cal A}_{\pm} \left( {\cal A}_{\pm}- 8\pi n^2 \right)}{32\pi {\cal M}({\cal A}_{\pm}-4\pi n^2)^2}  \\
\Phi_{\pm}^n &=& \frac{n \left(16\pi n^2 {\cal A}_{\pm}-\frac{3}{2} {{\cal A}_{\pm}}^2
-32\pi^2 n^4 \right)}{2{\cal M}({\cal A}_{\pm}-4\pi n^2)^2}
~. \label{invtn}
\end{eqnarray}
where,
\begin{eqnarray}
\Gamma_{\pm} &=& \mbox{Effective surface tension of ${\cal H}^{+}$ and ${\cal H}^{-}$} \nonumber \\
\Phi_{\pm}^n &=& \mbox{TN potential of ${\cal H}^\pm$ for NUT parameter} \nonumber
\end{eqnarray}

Using the Euler's theorem on homogenous function to ${\cal M}$ of degree $\frac{1}{2}$ in $({\cal A}_{\pm},n^2)$,
one can derive the mass for TN BH:
\begin{eqnarray}
{\cal M} &=& 2\Gamma_{\pm} {\cal A}_{\pm} +\Phi_{\pm}^{n} n
~. \label{bitn}
\end{eqnarray}
Remarkably, $\Gamma_{\pm}$ and $\Phi_{\pm}^n$ could be defined and are constant over the
${\cal H}^+$ and ${\cal H}^-$ for any stationary space-time.

Generally, combining the mass differential Eq. (\ref{dmtn}) with the first law leads to the Smarr-Gibbs-Duhem
relation. It is clearly evident from Eqs. (\ref{dmtn}) and (\ref{bitn}), such equations do not hold for
TN spacetime because
\begin{eqnarray}
\Gamma_{\pm} &=& \frac{\partial {\cal M}}{\partial {\cal A}_{\pm}} \neq \frac{{\kappa}_{\pm}} {8\pi}
\end{eqnarray}
An important point should be noted here that \emph{the first law of BH thermodynamics} does not
hold for TN spacetime. The reason behind that the Lorentzian TN spacetime
containing Dirac-Misner singularity, a coordinate type singularity
which may be considered as a manifestation of a non-trivial topological twisting
of the manifold \cite{aliev}. Which was first discovered by Misner in his paper in 1963.
Misner noticed that he could be remove this type of singularity by
using different type of time coordinates near the north and south pole.
However, this coordinate transformation made the time periodic. Thus for periodic time coordinate
we need Euclidean geometry. Since we are restricted here for Lorentzian geometry, this  could
be found  in elsewhere.

In case of RN BH, we may noted that the mass formula is given by
\begin{eqnarray}
{\cal M}^2 &=& \frac{{\cal A}_{\pm}}{16\pi}+\frac{\pi Q^4}{{\cal A}_{\pm}}+\frac{Q^2}{2} ~.\label{mrn}
\end{eqnarray}
Again the mass could be expressed in a simple bilinear form:
\begin{eqnarray}
{\cal M} &=& 2\Gamma_{\pm} {\cal A}_{\pm} +\Phi_{\pm} Q
~. \label{birn}
\end{eqnarray}
where,
\begin{eqnarray}
\Gamma_{\pm} &=& \frac{\partial {\cal M}}{\partial {\cal A}_{\pm}}
= \frac{{\kappa}_{\pm}} {8\pi} ~. \label{garn}
\end{eqnarray}
and
\begin{eqnarray}
\Phi_{\pm} &=& \frac{\partial {\cal M}}{\partial Q} =\frac{1}{{\cal M}}
\left(\frac{Q}{2}+\frac{2\pi Q^3}{{\cal A}_{\pm}}  \right)
 ~. \label{invrn}
\end{eqnarray}

Thus the Smarr-Gibbs-Duhem relation can be obtained as
\begin{eqnarray}
\frac{{\cal M}}{2} &=& {T}_{\pm}{\cal S}_{\pm}+\frac{\Phi_{\pm}Q}{2}
~. \label{birn1}
\end{eqnarray}
It can be seen that both Smarr-Gibbs-Duhem relation and First law of BH thermodynamics
satisfied in case of RN BH. In \cite{rabin}, the author derived the generalized Smarr formula for RN BH and 
KN BH for ${\cal H}^+$ only. We proved in \cite{pp14} this is valid for ${\cal H}^-$ also.

\subsection{ Christodoulou's Irreducible Mass for TN Spacetime:}

Christodoulou had shown that the irreducible mass of the BH, ${\cal M}_{irr}$
is proportional to the square root of the BH's surface area. He had emphasized
that the irreducible mass of a Kerr BH can be expressed  in terms of event
horizon area. We have already been suggested that for KN BH \cite{pp14} the irreducible mass
could be expressed in terms of both outer and inner horizon area. That means
the irreducible mass could be defined as
\begin{eqnarray}
 {\cal M}_{irr,\pm} &=& \sqrt{\frac{{\cal A}_{\pm}}{16\pi}}=\sqrt{\frac{{\cal S}_{\pm}}{16\pi}}
~. \label{irtn}
\end{eqnarray}
where,  $+$ indicates for ${\cal H}^+$ and $-$ indicates for ${\cal H}^-$.
Thus for TN BH this could be written as
\begin{eqnarray}
{\cal M}_{irr,\pm} &=& \frac{\sqrt{r_{\pm}^2+n^2}}{2}
~. \label{irtn1}
\end{eqnarray}
The product of the inner irreducible mass and outer irreducible mass  for TN
space-time is given by
\begin{eqnarray}
 {\cal M}_{irr,+} {\cal M}_{irr,-} &=& \sqrt{n^2({\cal M}^2+n^2)} ~. \label{irrmptn}
\end{eqnarray}
and therefore depends on the mass of the BH.

Another interesting formula i.e. the Christodoulou-Ruffini mass formula for TN
space-time in term of its irreducible mass and NUT parameter $n$ could be derived as
\begin{eqnarray}
{\cal M}^2 = \left[({\cal M}_{irr, \pm})^2-n^2\left(1-\frac{n^2}{4({\cal M}_{irr, \pm})^2}\right)\right]
\times \nonumber\\
\left( 1-\frac{n^2}{4 ({\cal M}_{irr, \pm})^2}\right)^{-1}  ~. \label{irrmtn}
\end{eqnarray}
When the NUT parameter  goes to zero we get the ${\cal M}={\cal M}_{irr}$ for Schwarzschild
space-time \cite{cr}. For our record, we may write the Christodoulou-Ruffini mass formula for 
RN BH:
\begin{eqnarray}
{\cal M} &=& {\cal M}_{irr,\pm}+\frac{Q^2}{4 {\cal M}_{irr,\pm} } ~.\label{irrmrn}
\end{eqnarray}

\section{\label{kerrt} Area Product and Area Sum for  KTN BH:}

Now we turn to the Kerr TN BH. It is  a stationary, axially
symmetric vacuum solution of Einstein equation with Kerr parameter $(a)$ and
NUT parameter $(n)$. When $n=0$, we obtain the Kerr geometry.

The metric \cite{demian,miller} is described  in Boyer-Lindquist
like spherical coordinates $(t, r, \theta, \phi)$:
$$
ds^2 = -\frac{\Delta}{\Sigma} \, \left[dt-\chi d\phi \right]^2+\frac{\sin^2\theta}{\Sigma}\,\left[(r^2+a^2+n^2)
\,d\phi-a dt\right]^2 +
$$
\begin{eqnarray}
\Sigma \, \left[\frac{dr^2}{\Delta}+d\theta^2\right] ~.\label{nktn}
\end{eqnarray}

where,
\begin{eqnarray}
a &\equiv&\frac{J}{{\cal M}},\, \Sigma \equiv r^2+(n+a\cos\theta)^2 \\
\Delta &\equiv& r^2-2{\cal M}r+a^2-n^2\\
\chi &\equiv& a\sin^2\theta-2n\cos\theta  ~.\label{adeltap}
\end{eqnarray}

Thus the metric is completely determined by the  three parameters i.e., the mass $({\cal M})$,
angular momentum ($J=a{\cal M}$) and  gravito-magnetic monopole or NUT parameter ($n$) or
magnetic mass. The spacetime has no curvature singularity, there are conical singularity
on its axis of symmetry  that result in the gravito-magnetic analogue of Dirac's string
quantization condition \cite{misner} and also leading to Dirac-Misner string singularity
at poles $\theta=0, \pi$. The conical singularity can be removed by imposing an
appropriate periodicity condition on the time coordinate. It should be noted that
the presence of the NUT parameter in the  spacetime destroy its asymptotic
structure making it asymptotically non-flat.

The radii of the horizons are given by
\begin{eqnarray}
r_{\pm}= {\cal M}\pm\sqrt{{\cal M}^2+n^2-a^2}
\end{eqnarray}

The area of inner and outer horizons are 
\begin{eqnarray}
{\cal A}_{\pm} &=& 8\pi\Big[({\cal M}^2+n^2) \pm {\cal M}\sqrt{{\cal M}^2+n^2-a^2} \Big]
~.\label{arKTN}
\end{eqnarray}
The product of ${\cal A}_{+} {\cal A}_{-}$ is given by
\begin{eqnarray}
{\cal A}_{+} {\cal A}_{-} &=& (8\pi)^2\Big[J^2+n^2({\cal M}^2+n^2)\Big] ~.\label{prktn}
\end{eqnarray}
and the sum of ${\cal A}_{+}+ {\cal A}_{-}$ is given by
\begin{eqnarray}
{\cal A}_{+}+ {\cal A}_{-} &=& 16 \pi\Big({\cal M}^2+n^2\Big) ~.\label{sumKNtn}
\end{eqnarray}
The area product and area sum  shown not to be universal for KTN BH.

The Bekenstein-Hawking \cite{bk1,bcw}  entropy associated with this BH reads
\begin{eqnarray}
{\cal S}_{\pm} &=& 2\pi \Big({\cal M}r_{\pm}+n^2 \Big)  ~.\label{etpktn}
\end{eqnarray}

Their product can be derived as
\begin{eqnarray}
{\cal S}_{+} {\cal S}_{-} &=& (2\pi)^2\Big[J^2+n^2({\cal M}^2+n^2)\Big] ~.\label{pentktn}
\end{eqnarray}
and their sum reads
\begin{eqnarray}
{\cal S}_{+}+ {\cal S}_{-} &=& 4\pi \Big({\cal M}^2+n^2\Big) ~.\label{entrtn}
\end{eqnarray}
It implies that entropy product and entropy sum both are depends on mass i.e. non-universal.

The surface gravity is at ${\cal H}^\pm$ given by
\begin{eqnarray}
{\kappa}_{\pm} &=& \frac{r_{\pm}-r_{\mp}}{ 2\left(2{\cal M}r_{\pm}+2n^2 \right)}  ~.\label{sgKtn}
\end{eqnarray}
and
the BH temperature or Hawking temperature of ${\cal H}^\pm$ reads as
\begin{eqnarray}
T_{\pm} &=& \frac{r_{\pm}-r_{\mp}}{4\pi (r_{\pm}^2+a^2+n^2)}  ~.\label{tmKtn}
\end{eqnarray}


\subsection{ The Mass Formula for  KTN Spacetime:}

Now it is straightforward calculation to compute the surface area of KTN BH
in terms of outer and inner horizon:
\begin{eqnarray}
{\cal A}_{\pm} &=& 8\pi \Big({\cal M}^2+n^2 \pm \sqrt{{\cal M}^4-J^2+n^2{\cal M}^2}\Big) ~.\label{arKtn}
\end{eqnarray}
It should be noted that the BH surface area is indeed constant over the ${\cal H}^{\pm}$.
Inverting the above equation which yields
\begin{eqnarray}
{\cal M}^2 &=&  \frac{1}{({\cal A}_{\pm}-4\pi n^2)} \left[\frac{{\cal A}_{\pm}^2}{16\pi}+4\pi J^2
-n^2({\cal A}_{\pm}-4\pi n^2)\right] ~.\label{massktn}
\end{eqnarray}
which shows that the BH mass can be expressed in terms of both the area of ${\cal H}^+$ and
${\cal H}^-$.

The mass differential could be derived in terms of three physical invariant of ${\cal H}^{\pm}$:
\begin{eqnarray}
d{\cal M} &=& \Gamma_{\pm} d{\cal A}_{\pm} + \Omega_{\pm} dJ +\Phi_{\pm}^{n}dn
~. \label{dmktn}
\end{eqnarray}
where
\begin{eqnarray}
\Gamma_{\pm} &=&  \frac{1}{2{\cal M}({\cal A}_{\pm}-4\pi n^2)^2} \left( \frac{{\cal A}_{\pm}^2}{16\pi}-4\pi J^2
-\frac{n^2{\cal A}_{\pm}}{2}  \right) \\
\Omega_{\pm} &=& \frac{4\pi J}{{\cal M}{(\cal A}_{\pm}-4\pi n^2)} \\
\Phi_{\pm}^n &=& \frac{\left(16\pi n^3 {\cal A}_{\pm}-\frac{3}{2}n {{\cal A}_{\pm}}^2
-32\pi^2 n^5+32n\pi J^2\right)}{2{\cal M}({\cal A}_{\pm}-4\pi n^2)^2}
~. \label{invktn}
\end{eqnarray}
where,
\begin{eqnarray}
\Gamma_{\pm} &=& \mbox{Effective surface tension of ${\cal H}^{+}$ and ${\cal H}^{-}$} \nonumber \\
\Omega_{\pm} &=&  \mbox{Angular velocity of ${\cal H}^\pm$} \nonumber \\
\Phi_{\pm}^n &=& \mbox{ TN potential of ${\cal H}^\pm$ for NUT charge} \nonumber
\end{eqnarray}
Since the mass of BH is homogenous of order $\frac{1}{2}$  in the variables
$({\cal A}_{\pm}, J, n^2)$, therefore using the Euler's theorem on homogenous function to ${\cal M}$ one obtains,
\begin{eqnarray}
{\cal M} &=& 2\Gamma_{\pm} {\cal A}_{\pm} + 2J \Omega_{\pm} +\Phi_{\pm}^{n} n ~. \label{biktn}
\end{eqnarray}
It implies that the ${\cal M}$ can be expressed in term of these quantities both for
${\cal H}^{\pm}$  as a  bi-linear form. It may be noted that $\Gamma_{\pm}$,  $\Omega_{\pm}$
and $\Phi_{\pm}^n$ could be defined and are indeed constant on the ${\cal H}^+$ and ${\cal H}^-$
for any stationary, axially symmetric spacetime.

Again the $d{\cal M}$ is a perfect differential, one can freely defined any path of
integration in $({\cal A}_{\pm}, J, n)$ space. Thus the surface energy
${\cal E}_{s, \pm}$ for ${\cal H}^{\pm}$  can be defined as
\begin{eqnarray}
{\cal E}_{s, \pm} &=& \int_{0}^{{\cal A}_{\pm}} \Gamma (\tilde{{\cal A}_{\pm}}
, 0 ,0) d\tilde{{\cal A}_{\pm}}; ~ \label{se}
\end{eqnarray}

The rotational energy  for ${\cal H}^{\pm}$ can be defined as
\begin{eqnarray}
{\cal E}_{r, \pm} &=& \int_{0}^{J} \Omega_{\pm} ({\cal A}_{\pm}
, \tilde{J} , 0) d\tilde{J},\,\,  \mbox{${\cal A}_{\pm}$ fixed}; ~ \label{re}
\end{eqnarray}
and finally  the electromagnetic  energy for ${\cal H}^{\pm}$ due to the
NUT charge $n$ is given by
\begin{eqnarray}
{\cal E}_{em, \pm} &=& \int_{0}^{n} \Phi_{\pm} ({\cal A}_{\pm}
, J, \tilde{n}) d\tilde{n},\,\, \mbox{${\cal A}_{\pm}$, $J$ fixed}; ~ \label{re2}
\end{eqnarray}

Now combining the mass differential Eq. (\ref{dmktn}), and the first law gives the Smarr-Gibbs-Duhem relation. We see
that from Eqs. (\ref{dmktn}) and (\ref{biktn}), such an equation does not hold for KTN BH because
\begin{eqnarray}
\Gamma_{\pm} &=& \frac{\partial {\cal M}}{\partial {\cal A}_{\pm}} \neq \frac{{\kappa}_{\pm}} {8\pi}=\frac{T_{\pm}}{4}
\end{eqnarray}
We also observed that \emph{the first law of BH thermodynamics} does not hold for KTN BH. The reason 
of failure is that due to the Dirac-Misner string singularity.

Whereas, the BH mass or ADM mass could be expressed in terms of area of both horizons ${\cal H}^\pm$:
\begin{eqnarray}
{\cal M}^2 &=& \frac{{\cal A}_{\pm}}{16\pi}+\frac{4\pi J^2}{{\cal A}_{\pm}} ~.\label{masskerr}
\end{eqnarray}
Again the mass could be expressed in terms of a simple bilinear form:
\begin{eqnarray}
{\cal M} &=& 2\Gamma_{\pm} {\cal A}_{\pm} + 2J \Omega_{\pm} ~. \label{bikerr}
\end{eqnarray}
where,
\begin{eqnarray}
\Gamma_{\pm} &=& \frac{\partial {\cal M}}{\partial {\cal A}_{\pm}} = \frac{{\kappa}_{\pm}} {8\pi} ~. \label{gakerr}
\end{eqnarray}
and
\begin{eqnarray}
\Omega_{\pm} &=& \frac{\partial {\cal M}}{\partial J}=\frac{4\pi J}{{\cal M}{\cal A}_{\pm}} ~. \label{invkerr}
\end{eqnarray}
The Smarr-Gibbs-Duhem relation holds for Kerr BH as
\begin{eqnarray}
\frac{{\cal M}}{2} &=& {T}_{\pm}{\cal S}_{\pm} + J\Omega_{\pm} ~. \label{biker}
\end{eqnarray}
It can be seen that both First law of BH thermodynamics and Smarr-Gibbs-Duhem relation
holds for Kerr BH.

\subsection{Christodoulou's Irreducible Mass for KTN Spacetime:}

Analogously, we can define the  irreducible mass for KTN BH:
\begin{eqnarray}
 {\cal M}_{irr, \pm} &=& \sqrt{\frac{{\cal A}_{\pm}}{16\pi}} =\frac{\sqrt{r_{\pm}^2+a^2+n^2}}{2}
~. \label{irktn}
\end{eqnarray}

Again, the area and angular velocity could be expressed in terms of ${\cal M}_{irr \pm}$:
\begin{eqnarray}
 {\cal A}_{\pm} &=& 16 \pi ({\cal M}_{irr,\pm})^2
~. \label{irktn1}
\end{eqnarray}
and
\begin{eqnarray}
 {\Omega}_{\pm} &=& \frac{a}{r_{\pm}^2+a^2+n^2} = \frac{a}{4({\cal M}_{irr,\pm})^2} ~. \label{iromktn}
\end{eqnarray}

Therefore the product of the irreducible mass of ${\cal H}^\pm$  for KTN
space-time:
\begin{eqnarray}
 {\cal M}_{irr,+} {\cal M}_{irr,-} &=&  \sqrt{J^2+n^2({\cal M}^2+n^2)} ~. \label{irrmpktn}
\end{eqnarray}
It is noteworthy that the product of irreducible mass of  ${\cal H}^\pm$ for KTN
BH, depends on mass and therefore is not \emph {universal}.

The Christodoulou-Ruffini mass formula for KTN spacetime in terms of
irreducible mass, angular momentum $J$ and NUT parameter $n$ is given by
\begin{eqnarray}
{\cal M}^2 = \Big[({\cal M}_{irr, \pm})^2+\frac{J^2}{4 ({\cal M}_{irr,\pm})^2}-n^2\Big(1-\frac{n^2}{4
({\cal M}_{irr, \pm})^2}\Big) \Big] \times \nonumber\\
\Big(1-\frac{n^2}{4 ({\cal M}_{irr,\pm})^2}\Big)^{-1} 
~. \label{irrmfktn}
\end{eqnarray}
When the NUT parameter goes to zero we get the mass formula for Kerr space-time:
\begin{eqnarray}
{\cal M}^2 = ({\cal M}_{irr, \pm})^2+\frac{J^2}{4 ({\cal M}_{irr,\pm})^2}
 ~. \label{irrkerr}
\end{eqnarray}

So far we have calculated different thermodynamic quantities and these formulae might be 
useful to further understanding the microscopic nature of BH entropy both exterior and interior. 
Again, the entropy products of inner and outer horizons could be used to determine whether the 
classical BH entropy could be written as a Cardy formula, giving some evidence for a holographic 
description of BH/CFT correspondence \cite{chen}. The above thermodynamic properties including the 
Hawking temperature and area of both the horizons may therefore be expected to play a crucial role 
in understanding the BH entropy of ${\cal H}^{\pm}$ at the microscopic level.

\section{\label{dis} Conclusion:}
In order to understand the BH entropy at the microscopic level, we have studied both the \emph{inner 
horizon and outer horizon thermodynamics} for TN and KTN BH in four dimensional Lorentzian geometry. 
We also computed various thermodynamic product and sums. Due to the presence 
of the NUT parameter, we have found that the thermodynamic products are not universal and have no nice 
quantization feature. It has been also shown that the first law of BH thermodynamics 
does not hold for these space-time. Moreover we  derived the mass formula and Christodoulou-Ruffini mass formula
of the above mentioned BHs. Furthermore we proved that Smarr-Gibbs-Duhem relation does not hold
for both the BH due to the presence of NUT parameter. We compared these properties with the 
RN BH and Kerr BH. It is shown that such features are unlikely in the RN BH and Kerr BH.

In \cite{wald}, the author showed the first law holds in an arbitrary theory of gravity derived from a 
diffeomorphism invariant Lagrangian and the quantity playing the role of the entropy of the BH was 
identified as the integral over the horizon of the `Noether charge' associated with the horizon Killing 
vector field. In this theory, the author has considered a certain set of assumptions. The main assumption
is the space-time should be asymptotic one. If one compare the results obtained here with the Wald's theory
one can get a little bit problem in fact it contradicts because in our case the space-time we considered is 
non-asymptotic one due to the NUT parameter, whereas the Wald's theory is applicable for only 
asymptotic space-time. 

The reason for the failure of the first law of BH thermodynamics and
Smarr-Gibbs-Duhem relation in the four dimensional Lorentzian geometry is
due to presence of the non-trivial NUT parameter. The another
reason is that Lorentzian TN and Lorentzian KTN geometry
contains the Misner-type singularity. A Misner string is a coordinate singularity
which may be considered as a manifestation of a non-trivial topological twisting
of the manifold.

In the orbit space $\Sigma$, the Misner string is a line such that the
integral of $A_{\mu}$ around a closed loop which encircles the string
does not vanish as the area of loop is taken to zero-it is the Dirac string
for the Kaluza-Klien vector potential $A_{\mu}$. The curvature is smooth along
the string, implying that it is a coordinate singularity.
Hawking and Hunter \cite{hh99} have shown that in $d$ dimensions, the entropy
could be expressed in terms of the $(d-2)$ obstructions to foliation, Bolts
and Misner strings by the universal formula:
\begin{eqnarray}
 S &=& \frac{{\cal A}_{Bolts}}{4}+\frac{{\cal A}_{MS}}{4}-\beta H_{MS}.  \nonumber
\end{eqnarray}
where ${\cal A}_{Bolts}$ and ${\cal A}_{MS}$ are respectively the $d-2$ volumes in the
Einstein frame of the Bolts and Mister strings and $H_{MS}$ is the Hamiltonian surface term
on the Misner strings. Also $\beta$ is the inverse of the Hawking temperature.
Thus it would be interesting to investigate the entropy product rules
due to the contribution from Dirac-Misner strings \cite{gibbons}.
This investigation might play a crucial role in the BH thermodynamics.
In summary, these thermodynamic product formulas of inner and outer horizons give us 
further understanding to the BH entropy (both interior and exterior) at the microscopic 
level.

\end{document}